\documentclass[twocolumn,showpacs,showkeys,preprintnumbers]{revtex4}

\usepackage{amssymb,amsmath,amsthm}
\usepackage{array,longtable}
\usepackage[dvips]{graphicx}
\usepackage[dvips]{color}
\newcommand{\version}{August 19, 2003}
\usepackage[mathscr]{eucal}
\usepackage[dvips]{graphicx}

\newcommand{\finkfile}{}

\renewcommand{\Im}{\mathop{\textmd{Im}}\nolimits}
\renewcommand{\Re}{\mathop{\textmd{Re}}\nolimits}

\newcolumntype{V}{>{$}m{4cm}<{$}}
\newcolumntype{C}{>{$}c<{$}}
\newcolumntype{L}{>{$}l<{$}}
\newcolumntype{R}{>{$}r<{$}}

\newcommand{\rx}{\textsl{x}}
\newcommand{\Ds}{\mathscr{D}}

\newcommand{\Zc}{\mathcal{Z}}
\newcommand{\Rh}{\mathbb R}

\newcommand{\ra}{\rightarrow}

\newcommand{\Ac}{\mathcal{A}}
\newcommand{\Nc}{\mathcal{N}}

\newcommand{\Hc}{\mathcal{H}}

\newcommand{\Mc}{\mathcal{M}}

\newcommand{\pd}{\partial}

\newcommand{\oz}{{\overline{z}}}
\newcommand{\ow}{{\overline{w}}}

\newcommand{\la}{\langle\!\langle}
\renewcommand{\ra}{\rangle\!\rangle}

\newcommand{\sgn}{\mathop{\mathrm{sgn}}\nolimits}

\renewcommand{\ap}{\alpha^{\prime}}

\allowdisplaybreaks[3]

\renewcommand{\theequation}{\arabic{section}.\arabic{equation}}

\begin{document}

\preprint{RUNHETC-2003-25}
\preprint{hep-th/0308147}

\title{Witten's Ghost Vertex Made Simple\\ ($bc$
and bosonized ghosts)}

\author{D.M.~Belov}
\altaffiliation[On leave from]{ Steklov Mathematical
Institute, Moscow, Russia}
\email{belov@physics.rutgers.edu}
\affiliation{
Department of Physics
\\
Rutgers University
\\
136 Frelinghuysen Rd., Piscataway, NJ 08854, USA
}

\date{\version}

\begin{abstract}
First, we diagonalize the $bc$-ghost
$3$-string Neumann matrices using the technique
described in hep-th/0304158.
Their eigenvalues
are in complete agreement with the previous authors.
Second, we diagonalize the $N$-string gluing vertices for
the bosonized ghost system. And third,  we verify  the descent
and associativity relations for the combined bosonic
matter+ghost gluing vertices.
We find that in order for these relations to be true, the vertices
must be normalized by the factor $\Zc_{N}$. Here $\Zc_{N}$
is the partition function of the bosonic
matter+ghost CFT on the gluing surface,
which is the unit disc with the Neumann boundary
conditions and
the midpoint cone like singularity specifying by
the angle excess $\pi(N-2)$.
\end{abstract}


\maketitle

\newpage

\tableofcontents

\section{Introduction}
\setcounter{equation}{0}
\dopage{\finkfile}

Witten's open cubic string field theory \cite{witten}
is usually formulated in terms of $N$-string gluing vertices
\cite{GJ,peskin2}. The expressions for
these vertices in the basis which diagonalizes $L_0$ lead to
complicated calculations.
Using the fact that $K_1=L_1+L_{-1}$ commutes with Witten's vertex,
Rastelli et al. \cite{spectroscopy} transformed it to the basis
with $K_1$ diagonal. They
found that the Neumann matrices in the zero momentum vertex take a simple
diagonal form in this basis. Further it was realized that
many calculations in the string field theory which looks
complicated in the $L_0$-basis can be done easily and analytically in the $K_1$-basis.
Using their technique \cite{spectroscopy} the succeeding authors  generalized
their result to including momenta \cite{Feng:2002rm,dima2}, ghosts
 \cite{dima2,Arefeva:2002jj,DK2}, background $B$-field \cite{Feng:2002ib}
and fermions \cite{Marino:2001ny}.
However all of these calculation were intrinsically indirect.
Recently the authors of \cite{Belov:2003df} formulated
a simple and direct method of changing the basis.
This method is so powerful that it allowed them to diagonalize
at once $N$-string Neumann matrices for all scale
dimensions in the matter sector,
$3$-string Neumann matrices for $\beta\gamma$ and $bc$ ghosts,
and to resolve the momentum difficulty.
However unlike all other cases the $3$-string $bc$-ghost's vertex was
diagonalized indirectly by relating it to the $6$-string
matter Neumann matrices \cite{GJ}.

The present paper has three aims. Firstly, we
diagonalize the $3$-string $bc$-ghost vertex
by directly changing the basis in its Neumann functions.

Secondly, we consider the general bosonized ghost
system \cite{fms} which is characterized by the
background charge $Q$ and the parity $\varepsilon=\pm 1$.
We find an expression for the $N$-string gluing
vertex of this system in the $K_1$-basis.
Then we show that ghost numbers
can be added to the vertex by a unitary
transformation, and discuss the differences of this
construction from the one for the matter sector \cite{Belov:2003df}.

Thirdly, we test the descent relations
\begin{equation}
{}_{1\dots N,N+1}\langle V_{N+1}|V_1\rangle_{N+1}
\stackrel{?}{=}
{}_{1\dots N}\,\langle V_{N}|
\label{ttdescent}
\end{equation}
and other associativity
relations for the combined bosonic matter+ghost gluing vertex $\langle V_N|$.
The need to verify them arises from several inconsistencies
in the calculations performed in the past two years.
First, there is a strange anomaly in the multiplication
of the wedge states $\langle h_N,0|$ \cite{problem} [Eq. (5.40) therein]. Second,
the direct calculation of the inner product
of two wedges $\langle h_3,3|h_3,0\rangle$
differs from the expected unity \cite{DK2,density}.
This result is in contradiction with the statement of
\cite{peskin2} [Eq. (5.59) therein].
And third, assuming that the descent
relation \eqref{ttdescent} are true for the vertices defined in \cite{GJ}
the authors of \cite{udi2}
find some contradictions in their calculations [compare Eqs. (3.34) and (3.38)
therein].
In the present paper we show that for critical
bosonic string there is a finite constant
$\Zc_{N+1,1;N}$
in the rhs of the descent relation \eqref{ttdescent}.
This constant can be written as
\begin{equation}
\Zc_{N+1,1;N}=\frac{\Zc_{N}}{\Zc_{1}\Zc_{N+1}},
\label{Zzzz}
\end{equation}
where the function $\Zc_{N}$ is
\begin{equation}
\Zc_{N}=\left[\frac{2}{N}\right]^{9/2}\,
\exp\biggl\{\frac{27}{2}\Bigl[D_0(N)-D_0(2)\Bigr]\biggr\}.
\end{equation}
The explicit expression for $D_0(N)$ is given in the text
of this paper. Here one only needs to know that
for $N\geqslant 1$ it monotonically increases, and goes to
zero as $1/N$. Now it is obvious that in order
to satisfy the descent relation one has
to normalize the vertices by
\begin{equation}
\langle V_N|\to\Zc_N\langle V_N|
\quad\text{for}\quad N\geqslant 1.
\label{normal}
\end{equation}
We also verify that the normalized vertices satisfy
all other associativity conditions. Notice that for $N=2$
the normalization $\Zc_N$ equals 1, therefore the string
inner product is not affected by \eqref{normal}. The function $\Zc_N$
is nothing else but the partition function of the matter+ghost CFT on the
gluing surface, which is the unit disk with Neumann
boundary conditions and midpoint cone like singularity
specifying by the angle excess $\pi(N-2)$.

For $N=3$ the vertex normalized as in Eq. \eqref{normal}
coincides with Witten's original definition \cite{witten}.
In that paper he defined it as the Polyakov integral over
the gluing surface. It seems that in most succeeding
papers the Polyakov integral was changed into the correlation function
on that surface, and the normalization factor $\Zc_N$ was lost.

\vspace{1cm}
The paper is organized as follows.
In Section~\ref{sec:not} we review the notations
used in \cite{Belov:2003df}. In Section~\ref{sec:ghost}
we diagonalize the $3$-string Neumann matrices for
the $bc$-ghost system. In Section~\ref{sec:NsB}
we consider the $N$-string gluing vertices for the general
bosonized ghost system.
We find its representation in the $K_1$ basis,
and describe how to change the ghost number by a unitary transformation.
In Section~\ref{sec:assoc} we prove the associativity
properties of the gluing vertices for the
combined bosonic matter+ghost system.
In Section~\ref{sec:disc} we discuss the influence
of the vertex normalization \eqref{normal} on numeric
calculations in SFT.
Appendix~\ref{app:st} contains necessary technical
information.

\section{Notations}
\label{sec:not}
\dopage{\finkfile}
\setcounter{equation}{0}
In this section we review the notations and
main formulae from \cite{Belov:2003df}.

Consider the primary discrete series $\Ds_s^+$ of the $SL(2,\Rh)$
representations. Here $s$ is the scale dimension, $s=0,\frac{1}{2},1,\dots$.
For example, $s=1$ corresponds to the zero momentum bosonic matter,
$s=\frac12$ to the fermions and $s=0$ to the bosonic matter with the zero modes.
An appropriate Hilbert space $\Hc_s$ consists of the functions $f(z)$
analytic inside the unit disk and square-integrable on the
boundary. The inner product is \cite{ruhl}
\begin{equation}
\langle g|f\rangle=\frac{1}{\pi\Gamma(2s-1)}\,
\int_{|z|\leqslant 1}\!\!d^2z\,\bigl[1-z\oz\bigr]^{2s-2}
\,\overline{g(z)}f(z).
\label{inner}
\end{equation}
The $s=\frac12$ singularity
is spurious \cite{ruhl},
but there is a real one as $s$ approaches zero.
The algebra $sl(2,\Rh)$ is generated by $L_0$, $L_{\pm 1}$
which are defined by
\begin{equation}
L_n=z^{n+1}\frac{d}{dz}+(n+1)\,sz^n.
\end{equation}
This representation is unitary for $s>0$.

\subsection{The discrete basis}
The elliptic generator $L_0$ has discrete eigenvalues $(m+s)$,
$m=0,1,2,\dots$. Its eigenfunctions normalized by \eqref{inner} are
\begin{multline}
|m,s\rangle(z)=N_m^{(s)}z^m
\\
\text{with}\quad
N_m^{(s)}=\biggl[\frac{\Gamma(m+2s)}{\Gamma(m+1)}\biggr]^{1/2}.
\label{basisn}
\end{multline}
Notice that for $s=0$ the only singular vector is $|0,0\rangle(z)$.

\subsection{The continuous basis}
The generator $K_1=L_1+L_{-1}$ commutes with Witten's star product
\cite{witten}, which therefore becomes simpler when it is
diagonalized \cite{spectroscopy}. It is convenient
to map
\begin{equation}
z=i\tanh w,
\label{zw}
\end{equation}
which takes the unit disk into the strip $|\Im w|\leqslant \frac{\pi}{4}$.
We assume that under a map $z\mapsto w$
the vector $f(z)$ transforms
in a trivial way
\begin{equation}
f(z)\mapsto f(z(w)).
\tag{\ref{zw}${}^{\prime}$}
\label{f->z}
\end{equation}
Then
\begin{equation}
K_1=-i\frac{d}{dw}+2is\tanh w.
\label{K_1}
\end{equation}
Since this is a hyperbolic generator, its
eigenvalues are all real numbers $\kappa$. The normalized eigenfunctions
of \eqref{K_1} are
\begin{equation}
|\kappa,s\rangle(z)=\bigl[A_s(\kappa)\bigr]^{1/2}\,
(\cosh w)^{2s}\,e^{i\kappa w},
\label{kappas}
\end{equation}
where $A_s(\kappa)$ is the normalization constant:
\begin{equation}
A_s(\kappa)=\frac{2^{2s-2}}{\pi}\,
\Gamma\Bigl(s+\frac{i\kappa}{2}\Bigr)
\Gamma\Bigl(s-\frac{i\kappa}{2}\Bigr).
\label{As}
\end{equation}

One sees that as $s\to0$ the function \eqref{As} becomes ill
defined at $\kappa=0$. Nevertheless the $s=0$ $K_1$-eigenfunctions
are well defined \cite{Belov:2003df}:
\begin{multline}
|\kappa,0\rangle(z)=\mathscr{P}\,\frac{\sqrt{A_1(\kappa)}}{\kappa}\,
e^{i\kappa w}
\\
=\mathscr{P}\,\frac{\sqrt{A_1(\kappa)}}{\kappa}+
|\kappa,\Omega\rangle(z).
\label{kappa0}
\end{multline}
Here $\mathscr{P}$ means the principal value, and
the function $|\kappa,\Omega\rangle(z)$ can also be written as
the integral of the $s=1$ $K_1$-eigenfunction
\begin{equation}
|\kappa,\Omega\rangle(z)\equiv
\int_0^{z}d\zeta\,|\kappa,1\rangle(\zeta).
\label{01rel}
\end{equation}
The vector $|\kappa,\Omega\rangle(z)$
is that which was found by Rastelli et al. \cite{spectroscopy}.
The important identity with $|\kappa,\Omega\rangle(z)$ is \cite{Belov:2003df}
\begin{subequations}
\begin{equation}
\frac{1}{2}\,\log(1+z^2)=\int_{-\infty}^{\infty}d\kappa\,
\mathscr{P}\frac{\sqrt{A_1(\kappa)}}{\kappa}\,|\kappa,\Omega\rangle(z).
\label{mid}
\end{equation}
Differentiating this with respect to $z$ and using \eqref{01rel}
one obtains another useful identity
\begin{equation}
\frac{z}{1+z^2}=\int_{-\infty}^{\infty}d\kappa\,
\mathscr{P}\frac{\sqrt{A_1(\kappa)}}{\kappa}\,|\kappa,1\rangle(z).
\label{mid2}
\end{equation}
\end{subequations}

We will frequently use the notation $\langle \kappa,s|(\oz)\equiv
\overline{|\kappa,s\rangle(z)}$.

\subsection{The transition matrix}
The transition matrix between the discrete and continuous bases is
an orthogonal matrix with elements
\begin{equation}
\langle m,s|\kappa,s\rangle=
V_m^{(s)}(\kappa)\,\frac{[A_s(\kappa)]^{\frac{1}{2}}}{N_m^{(s)}}.
\label{<mk>}
\end{equation}
Here the polynomials $V_m^{(s)}(\kappa)$ are given by
the generating function
\begin{equation}
(\cosh w)^{2s}\,e^{i\kappa w}
=\sum_{m=0}^{\infty} V_m^{(s)}(\kappa)z^m.
\label{polyV}
\end{equation}

Due to equation \eqref{01rel} the transition matrices
for $s=0$ and $s=1$ are related as \cite{Belov:2003df}
\begin{equation}
\langle m+1,0|\kappa,0\rangle=\langle m,1|\kappa,1\rangle
\qquad (m\geqslant 0),
\label{0=1}
\end{equation}
so $s=1$ is just $s=0$ with $m=0$ omitted.

\section{$3$-string Vertex for $bc$-system}
\label{sec:ghost}
\setcounter{equation}{0}
\dopage{\finkfile}
\subsection{Overview}
The $bc$-ghost system has a background charge $Q=-3$.
Due to conservation of this charge it is convenient
to write a $3$-string vertex over the vacuum
$\langle +|$, which is the conjugate of the ghost number $1$ vacuum $|-\rangle$
(i.e. $\langle +|-\rangle=1$).
These vacua are defined by
\begin{subequations}
\begin{alignat}{4}
b_n|-\rangle &=0\;\, (n>-1); &\quad
\langle +|b_n &=0\;\, (n< 0);
\\
c_m|-\rangle &=0\;\, (m\geqslant 1);
&\quad
\langle +|c_m &=0\;\,(m\leqslant 0);
\end{alignat}
\end{subequations}
so $\{b_{0},\,b_{1},\dots\}$ and $\{c_{1},\,c_{2},\dots\}$
are the annihilation operators.
The vacuum $|-\rangle$ is
related to the $SL(2,\Rh)$ invariant vacuum
$|0\rangle$ by $|-\rangle=c(0)|0\rangle$.
The $3$-string vertex over these vacua was constructed in \cite{GJ} [paper II],
and in our notations it reads
\begin{multline}
\langle V_3|=N_{bc}\,{}_{123}\langle +|
\\
\times\exp\left[-
\sum_{m=0,n=1}^{\infty} b_m^{(I)}\Bigl(\Mc^{IJ}_{bc}
C\Bigr)_{mn} c_n^{(J)}
\right]
\label{V3osc}
\end{multline}
Here ${}_{123}\langle +|$ means the tensor
product of three Fock vacua $\langle +|$,
$\bigl(\Mc^{IJ}_{bc}\bigr)_{mn}$ are the $3$-string
ghost Neumann matrices and $C_{mn}=(-1)^n\delta_{mn}$.

To obtain an expression for the matrix elements
$\bigl(\Mc^{IJ}_{bc}\bigr)_{mn}$ one can
calculate  the function
\begin{equation}
\langle V_3| c^{(I)}(z) b^{(J)}(z') |-\rangle_{123}
\end{equation}
in two different ways: using expression \eqref{V3osc},
and using the conformal definition of the vertex.
The details of this calculation can be found,  for example, in
Section~4 of \cite{Bonora:2003xp} or in \cite{jevicki}.
In our notations the result is
\begin{multline}
\Mc^{IJ}_{bc}(z,\oz')\equiv\sum_{m,n=0}^{\infty}
\bigl(\Mc^{IJ}_{bc}\bigr)_{n,m+1} z^n \oz^{\prime\,m}
\\
=\frac{\bigl[h_I'(z)\bigr]^{-1}\bigl[h_J'(-\oz')\bigr]^{2}}{h_I(z)-h_J(-\oz')}\,
\frac{\bigl[h_I(z)\bigr]^3-1}{\bigl[h_J(-\oz')\bigr]^3-1}
\left[\frac{\oz'}{z}\right]
\\
+
\frac{\delta^{IJ}}{z+\oz'},
\label{Mgh}
\end{multline}
and $N_{bc}=\left[\frac{3\sqrt{3}}{4}\right]^3$.
Here the maps $h_I(z)$ are
\begin{subequations}
\begin{align}
h_I(z)&=e^{i\varphi_I}\left(\frac{1-iz}{1+iz}\right)^{2/3}
=e^{i\varphi_I}\,e^{\frac{4w}{3}},
\\
h_I'(z)&=-\frac{4i}{3}\,\bigl(\cosh w\bigr)^2 h_I(z),
\end{align}
\label{maps}
\end{subequations}
where $z=i\tanh w$, and $\varphi_I=\frac{2\pi}{3}(2-I)$ for $I=1,2,3$.

\subsection{Diagonalizing Witten's $3$-string ghost vertex}
\label{sec:W3g}
The aim of this section is to rewrite the operator $\Mc_{bc}^{IJ}(z,\oz')$
in the $K_1$-basis. It is known
that the vertices commute with the operator $K_1$ \cite{GJ}.
So one expects that $\Mc_{bc}^{IJ}(z,\oz')$
takes a simple form in the $K_1$-basis.

\begin{figure}[!t]
\centering
\includegraphics[width=210pt]{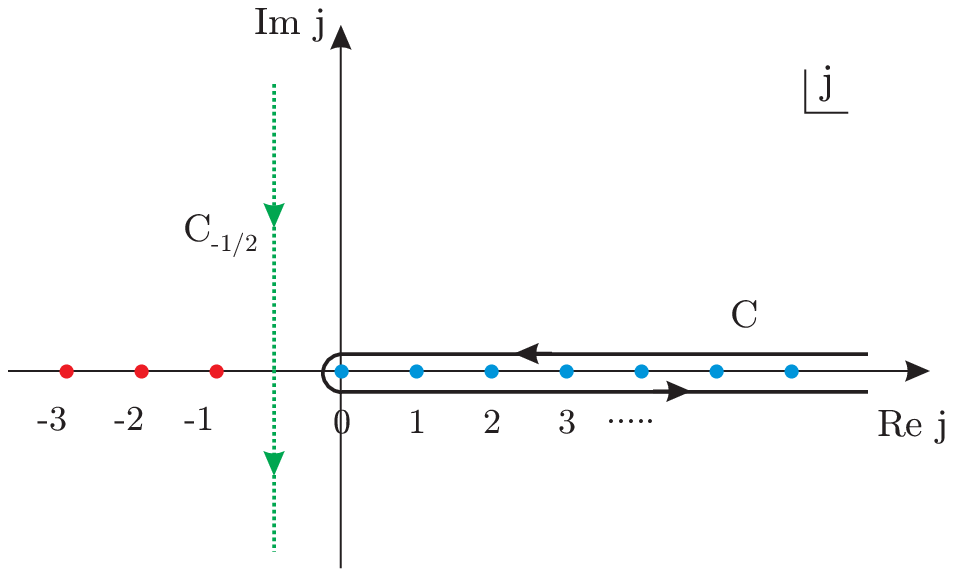}
\caption{The dots represent the poles
of the integrand in \eqref{Mgh1}. Contour
$C$ encircles the positive real
axis counterclockwise. Then we deform it to
the contour $C_{-\frac12}$, which lies parallel to the
imaginary axis at $\Re j =-\frac12$.}
\label{fig:1}
\end{figure}

To diagonalize the Neumann matrix \eqref{Mgh} we
first notice that the strange factors in \eqref{Mgh}
have very simple expression in the variable $w$
\begin{subequations}
\begin{align}
\frac{\bigl[h_I(z)\bigr]^3-1}{z}&=-4i\, e^{2w}(\cosh w)^2
\intertext{and}
\frac{\bigl[h_J(-\oz')\bigr]^3-1}{\oz'}&=+4i\, e^{2\ow'}(\cosh \ow')^2.
\end{align}
\label{p1}
\end{subequations}
Now we proceed as in \cite{Belov:2003df} ---
first do a binomial expansion of \eqref{Mgh}, then
rewrite it as a contour integral, and
finally do a Watson-Sommerfeld transformation.
Assuming $\Re (\ow'-w)<0$ we obtain
\begin{multline}
\Mc_{bc}^{IJ}(z,\oz')=\frac{4i}{3}\,\bigl(\cosh \ow'\bigr)^2\,
(-1)^{I-J}
\\
\times
\oint_C \frac{dj}{2i\sin \pi j}\,e^{\pm i\pi j}\,\left[e^{i(\varphi_J-\varphi_I)}
e^{\frac{4}{3}(\ow'-w)}\right]^{j+\frac12}
\\
+\frac{\delta^{IJ}}{z+\oz'}\label{Mgh1}
\end{multline}
where the contour $C$ encircles the positive real axis counterclockwise
(see Figure~\ref{fig:1}). Notice that we have a
sign ambiguity in the exponential, which comes
from the analytic continuation of $(-1)^j$.

Before deforming the contour as in Figure~\ref{fig:1} we must
worry about the convergence at infinity. Starting from
here we will consider the cases $I\ne J$ and $I=J$ separately.

\subsection{Matrices $\Mc_{bc}^{IJ}$ for $I\ne J$}

For $I\ne J$ Eq. \eqref{Mgh1} can be rewritten as
\begin{multline}
\Mc_{bc}^{IJ}(z,\oz')=(\mp 1)\frac{4i}{3}\,\bigl(\cosh \ow'\bigr)^2\,
(-1)^{I-J}
\\
\times\oint_C \frac{dj}{2\sin(\pi j)}\left[e^{i(\varphi_J-\varphi_I\pm \pi)}
e^{\frac{4}{3}(\ow'-w)}\right]^{j+\frac12}.
\label{MghIneJ0}
\end{multline}
To deform  the contour as it is shown in Figure~\ref{fig:1}
we must worry about convergence as $|\Im j|\to\infty$.
To this end for $I<J$ we choose the upper ``$+$'' sign
in the exponential. This guarantees that for $1\leqslant I<J\leqslant 3$
\begin{equation*}
|\varphi_J-\varphi_I+\pi|\leqslant \frac{\pi}{3}.
\end{equation*}
Taking into account the following asymptotics as $|\Im j|\to\infty$
\begin{align*}
&\frac{1}{\sin(\pi j)}\propto e^{-\pi|\Im j|}
\intertext{and}
&\left|e^{\frac{4}{3}(\ow'-w)\,j}\right|\leqslant e^{\frac{4}{3}|\Im(\ow'-w)||\Im j|},
\end{align*}
one concludes that for arbitrary small $\delta>0$
and $|\Im(\ow'-w)|\leqslant \frac{\pi}{2}-\delta$
the integrand has at least the exponential falloff $e^{-\delta|\Im j|}$.
(Oppositely, for $J<I$ one
has to choose the lower ``$-$'' sign in the exponential. This
guarantees the same falloff as $|\Im j|\to \infty$).
Now we can shift the contour $C$ to $\Re j=-\frac12$ by writing
\begin{equation}
j=-\frac12-\frac{3i\kappa}{4}
\label{jk}
\end{equation}
to get (for $I< J$)
\begin{multline}
\Mc_{bc}^{IJ}(z,\oz')=(-1)^{I-J}\bigl(\cosh \ow'\bigr)^2\,
\\
\times\int_{-\infty}^{\infty}d\kappa\,\frac{e^{\frac{\pi\kappa}{4}(2I-2J+3)}}{2\cosh\frac{3\pi\kappa}{4}}\,
e^{i\kappa(w-\ow')}.
\label{MghIneJ1}
\end{multline}
Notice that the integral here converges now for all $w,\ow'$ in
the strip $|\Im w|<\frac{\pi}{4}$. Therefore (by the
standard analytic continuation
arguments) the right hand
side represents the operator \eqref{Mgh} for all
$z$ and $\oz'$ in the unit disk.

Comparing the $\cosh$-factors
in Eq. \eqref{MghIneJ1} with the ones in Eq. \eqref{kappas}
one concludes that the continuum
$K_1$-eigenfunctions correspond to $s=0$ and $s=1$.
From Eqs. \eqref{kappa0} and \eqref{kappas} it follows
that the normalization factor
of their tensor product must be $\mathscr{P}\,\frac{A_1(\kappa)}{\kappa}$.
Insertion of unity $1=\frac{\kappa}{A_1(\kappa)}\,
\mathscr{P}\,\frac{A_1(\kappa)}{\kappa}$ into Eq. \eqref{MghIneJ1}
yields
\begin{subequations}
\begin{equation}
\Mc_{bc}^{IJ}(z,\oz')=
-\int_{-\infty}^{\infty}d\kappa\,\mu^{IJ}_{bc}(\kappa)\,
|\kappa,0\rangle(z)\otimes\langle\kappa,1|(\oz'),
\end{equation}
where
\begin{align}
\mu^{12}_{bc}(\kappa)&=+
e^{+\rx}\,\frac{\sinh 2\rx}{\cosh 3\rx},
\\
\mu^{13}_{bc}(\kappa)&=-
e^{-\rx}\,
\frac{\sinh 2\rx}{\cosh 3\rx}
\end{align}
\label{MIneJ1}
\end{subequations}
with $\rx\equiv\frac{\pi\kappa}{4}$.

\subsection{Matrix $\Mc^{II}_{bc}$}

For $I=J$ expression \eqref{Mgh} becomes
\begin{multline}
\Mc_{bc}^{II}(z,\oz')=\frac{1}{z+\oz'}+\frac{4i}{3}\,\bigl(\cosh \ow'\bigr)^2
\\
\times\oint_C \frac{dj}{2i \sin(\pi j)}\,e^{\pm i\pi j}\,
\left[e^{\frac{4}{3}(\ow'-w)}\right]^{j+\frac12}
.
\label{MII}
\end{multline}
This expression contains two terms, and we will consider them
separately.
In the second term we want to deform the contour $C$ as in Figure~\ref{fig:1}.
To this end we must first  worry about convergence at infinity.
For general $w$ and $\ow'$ the integrand does not go
to zero as $|\Im j|\to \infty$. However for
$0< \mp \Im (\ow'-w)\leqslant\frac{\pi}{2}$ the integrand has
an exponential falloff. In this case we deform the contour
$C$ as in Figure~\ref{fig:1}
by writing \eqref{jk} to get
\begin{multline*}
\text{second term}=
\\
\pm\bigl(\cosh \ow'\bigr)^2\,
\int_{-\infty}^{\infty}d\kappa\,
\frac{e^{\mp\frac{3\pi\kappa}{4}}}{2\cosh \frac{3\pi\kappa}{4}}
\,e^{i\kappa(w-\ow')}.
\end{multline*}
Comparing the $\cosh$-factors here
with the ones in \eqref{kappas} one concludes that the continuum
eigenfunctions correspond to $s=0$ and $s=1$.
Therefore, we can write
\begin{multline}
\text{second term}=\pm
\int_{-\infty}^{\infty}d\kappa\,
e^{\pm\frac{3\pi\kappa}{4}}\,\frac{\sinh\frac{\pi\kappa}{2}}{\cosh \frac{3\pi\kappa}{4}}
\\
\times
|\kappa,0\rangle(z)\otimes\langle \kappa,1|(\oz'),
\label{II-1}
\end{multline}
where the eigenfunctions
$|\kappa,0\rangle$ and $\langle\kappa,1|$
are defined in Eqs. \eqref{kappa0} and \eqref{kappas}
respectively.
The representation \eqref{II-1} is valid only
for $0< \mp \Im (\ow'-w)\leqslant\frac{\pi}{2}$.

Now we need to represent the first term in Eq. \eqref{MII} through
the tensor product of $s=0$ and $s=1$ eigenfunctions.
The details of this calculation can be found in Appendix~\ref{app:st}.
The result is
\begin{multline}
\frac{1}{z+\oz'}=-\int_{-\infty}^{\infty}d\kappa\,
e^{\pm\frac{\pi\kappa}{2}}\,|\kappa,0\rangle(z)\otimes \langle\kappa,1|(\oz')
\\
+\int_{-\infty}^{\infty}d\kappa\,
\mathscr{P}\,\frac{\sqrt{A_1(\kappa)}}{\kappa}\otimes
\langle\kappa,1|(\oz').
\label{II-2}
\end{multline}
The first integral here converges for $0<\mp\Im(\ow'-w)\leqslant\frac{\pi}{2}$,
while the second integral converges for all $\oz'$ in the unit disk.

Substitution of Eqs. \eqref{II-1} and \eqref{II-2} into \eqref{MII} yields
\begin{multline}
\Mc_{bc}^{II}(z,\oz')=
\int_{-\infty}^{\infty}d\kappa\,\mathscr{P}\,\frac{\sqrt{A_1(\kappa)}}{\kappa}\otimes
\langle\kappa,1|(\oz')
\\
+\int_{-\infty}^{\infty}d\kappa\,
\left[\pm e^{\pm\frac{3\pi}{4}}\,\frac{\sinh\frac{\pi\kappa}{2}}{\cosh \frac{3\pi\kappa}{4}}
-e^{\pm\frac{\pi\kappa}{2}}
\right]
\\
\times
|\kappa,0\rangle(z)\otimes\langle \kappa,1|(\oz')
.
\end{multline}
This expression has a sign ambiguity in the second term.
However if we want the first integral
to be convergent in the region $|\Im (w-\ow')|\leqslant \frac{\pi}{2}$
we must choose the upper ``$+$'' sign in this expression.
Finally, we get
\begin{subequations}
\begin{multline}
\Mc_{bc}^{II}(z,\oz')=
-\int_{-\infty}^{\infty}d\kappa\,
\mu_{bc}^{II}(\kappa)\,
|\kappa,0\rangle(z)\otimes\langle \kappa,1|(\oz')
\\
+\int_{-\infty}^{\infty}d\kappa\,\mathscr{P}\,\frac{\sqrt{A_1(\kappa)}}{\kappa}\otimes
\langle\kappa,1|(\oz'),
\end{multline}
where
\begin{equation}
\mu_{bc}^{II}(\kappa)=\frac{\cosh \rx}{\cosh 3\rx}\quad
\text{with}\quad \rx\equiv\frac{\pi\kappa}{4}.
\end{equation}
\label{MII1}
\end{subequations}

\subsection{The Neumann matrices in the discrete basis}
Collecting equations \eqref{MII1} and \eqref{MIneJ1}, one
concludes that the ghost $3$-string Neumann functions
\eqref{Mgh}
have the following diagonal representation in the $K_1$-basis
\begin{subequations}
\begin{multline}
\Mc_{bc}^{IJ}(z,\oz')=
-\int_{-\infty}^{\infty}d\kappa\,
\mu_{bc}^{IJ}(\kappa)\,
|\kappa,0\rangle(z)\otimes\langle \kappa,1|(\oz')
\\
+\delta^{IJ}\,\int_{-\infty}^{\infty}d\kappa
\,\mathscr{P}\,\frac{\sqrt{A_1(\kappa)}}{\kappa}\otimes
\langle\kappa,1|(\oz'),
\end{multline}
where $1\leqslant I,J\leqslant 3$,
\begin{align}
\mu_{bc}^{II}(\kappa)&=\frac{\cosh \rx}{\cosh 3\rx},
\\
\mu_{bc}^{I,I+1}(\kappa)&=+e^{\rx}\,\frac{\sinh 2\rx}{\cosh 3\rx},
\\
\mu_{bc}^{I+1,I}(\kappa)&=-e^{-\rx}\,\frac{\sinh 2\rx}{\cosh 3\rx}
\label{Mbceigen}
\end{align}
and $\rx\equiv \frac{\pi\kappa}{4}$.
\end{subequations}

To find the Neumann matrices $\bigl(\Mc_{bc}^{IJ}\bigr)_{n,m+1}$
we can calculate the matrix elements of the operator
$\Mc_{bc}^{IJ}(z,\oz')$ between the vectors
$\langle n,0|$ and $|m,1\rangle$. In this calculation
one obviously gets a divergence for $n=0$ (see Eq. \eqref{basisn}).
We can handle this divergence
by considering instead of the operator $\Mc_{bc}^{IJ}(z,\oz')$
the vector $\Mc_{bc}^{IJ}(0,\oz')$ and calculate
its inner product with $|m,1\rangle$.
Proceeding in this way one obtains
\begin{subequations}
\begin{multline}
\bigl(\Mc_{bc}^{IJ}\bigr)_{n+1,m+1}
=
-\frac{\sqrt{m+1}}{\sqrt{n+1}}\,\int_{-\infty}^{\infty}d\kappa\,\mu^{IJ}_{bc}(\kappa)
\\
\times
\langle n+1,0|\kappa,0\rangle\,\langle\kappa,1|m,1\rangle
\end{multline}
for $n,m\geqslant 0$ and
\begin{multline}
\bigl(\Mc_{bc}^{IJ}\bigr)_{0,m+1}=
-\sqrt{m+1}\,\int_{-\infty}^{\infty}d\kappa\,\Bigl[\mu^{IJ}_{bc}(\kappa)
-\delta^{IJ}\Bigr]
\\
\times
\frac{\sqrt{A_1(\kappa)}}{\kappa}\,\langle\kappa,1|m,1\rangle.
\end{multline}
\label{MghD}
\end{subequations}
Here the square roots come from calculation of the inner
products of $z^n$ or $\oz^{\prime m}$,
appearing in the definition \eqref{Mgh} of $\Mc_{bc}^{IJ}$,
with the vectors $\langle n,0|$ or $|m,1\rangle$.

Taking into account Eq. \eqref{0=1} and $\mu^{IJ}_{bc}(0)=\delta^{IJ}$
we find that the representation \eqref{MghD}
and eigenvalues \eqref{Mbceigen}
completely agree with the ones obtained in
\cite{Belov:2003df} [Eqs. (6.7) and (6.8) therein].

\section{$N$-string Vertex for Bosonized Ghosts}
\label{sec:NsB}
\setcounter{equation}{0}

\subsection{Overview}
The aim of this subsection is to review a construction of
LeClair et al. \cite{peskin2} of the $N$-string
gluing vertex for the bosonized ghosts.

Here we consider CFT for the general bosonized ghost system \cite{fms},
which is characterized by the background charge $Q$
and the parity $\varepsilon=\pm 1$.
The ghost number current $j(z)=\varepsilon \pd\phi(z)$
is an anomalous primary operator of dimension $1$, and transforms
under a conformal map $h(z)$ as
\begin{equation}
j(z)\mapsto \bigl(h\circ j\bigr)(z)=h'(z)j(h(z))
+\frac{Q}{2}\,\frac{h''(z)}{h'(z)}.
\label{h(j)}
\end{equation}
This current has the following mode expansion
\begin{multline}
j(z)=\frac{j_0}{z}+\sum_{n=1}^{\infty}\sqrt{n}\,
\Bigl\{a_n^+ z^{n-1}-a_n^- z^{-n-1}\Bigr\}
\\
\text{with}\quad
[a_n^-,\,a_m^+]=-\varepsilon\, \delta_{mn}.
\end{multline}
Here $a^{\pm}_m$ are creation/annihilation operators
over the vacuum $|q\rangle$, which is an eigenvector of the operator $j_0$
with eigenvalue $q$. Due to the anomalous transformation
law \eqref{h(j)} the conjugate vacuum to $|q\rangle$ is $\langle -q-Q|$.
From this it follows that $j^{\dag}_0=-j_0-Q$.

The OPEs of the fields $\phi(z)$ and $j(z)$ are
\begin{subequations}
\begin{align}
&\phi(z)\phi(z') \sim \varepsilon \log(z-z'),
\\
&j(z)\,e^{q\phi(z)}\sim\frac{q}{z-z'}\,e^{q\phi(z')}.
\end{align}
\end{subequations}
The matter field $X^{\mu}$ can be obtained from the expression above
by identifying $\varepsilon$ with $-\eta^{\mu\nu}$,
$j_0=-i\sqrt{2\ap}\,p^{\mu}$
and $a^{\pm}_m=\mp\frac{i}{\sqrt{m}}\,\alpha_{\mp m}^{\mu}$.

\begin{widetext}
The gluing vertex for the bosonized ghosts
differs in two ways from that for the $X$ field.
First, it has nonzero background charge $Q$,
and second the momentum eigenvalues are no longer continuous
but form a discrete set. The vertex reads \cite{peskin2} [Eq. (5.1) therein]
\begin{multline}
\langle V_{N,\,Q}^{(0)}|=
\sum_{\{q^I\}}\delta_{q^1+\dots+q^N+Q,0}\,
\bigotimes_{I=1}^N{}_I\langle-q^I-Q|
\,\exp\Biggl[
\frac{\varepsilon}{2}\,\sum_{I=1}^N q^I(Q+q^I)
+\frac{\varepsilon}{2}\,\sum_{I\ne J}q^I N_{00}^{IJ}q^J
\\
-\frac{\varepsilon}{2}\sum_{I=1}^N\sum_{n=0}^{\infty}
Q\,K_n^Ia_n^{-(I)}
-\varepsilon\,\sum_{I,J=1}^N\sum_{n=1}^{\infty}q^I N_{0n}^{IJ}a_n^{-(J)}
+\frac{\varepsilon}{2}\,\sum_{I,J=1}^N\sum_{m,n=1}^{\infty}
a_m^{-(I)}N_{mn}^{IJ}a_n^{-(J)}
\Biggr],
\end{multline}
where the Neumann function coefficients $N^{IJ}_{nm}$
are defined by the same formula as the ones for the matter part of the vertex
\cite{peskin2}
and
\begin{equation}
K^I(z)\equiv\sum_{n=1}^{\infty}K_{n}^I\,\sqrt{n}\,z^n
=\frac{h''_I(z)}{h_I'(z)}.
\end{equation}
The fact that the terms in the exponent contain the coefficient $Q$
is a direct consequence of the transformation law \eqref{h(j)}.
Using the \textit{anomalous} momentum conservation law and
$\langle -q-Q|j_0=\langle -q-Q|q$ we can rewrite the exponential as \cite{peskin2}
\begin{equation}
\exp\Biggl[
\frac{\varepsilon}{2}\,\sum_{I,J}j_0^I\Nc_{00}^{IJ}j_0^J
-\varepsilon\,\sum_{I,J}\sum_{n=1}^{\infty}j_0^I \Nc_{0n}^{IJ}a_n^{-(J)}
+\frac{\varepsilon}{2}\,\sum_{I,J}\sum_{m,n=1}^{\infty}
a_m^{-(I)}\Nc_{mn}^{IJ}a_n^{-(J)}
\Biggr],
\label{expNc}
\end{equation}
\end{widetext}
where the new Neumann-function coefficients are related to the old
ones by
\begin{subequations}
\begin{align}
\Nc_{00}^{IJ}&=N_{00}^{IJ}-\frac12\,N_{00}^{II}-\frac{1}{2}\,N_{00}^{JJ},
\\
\Nc_{0m}^{IJ}&=N_{0m}^{IJ}-\frac{1}{2}\,K_{m}^{J},
\\
\Nc_{mn}^{IJ}&=N_{mn}^{IJ}.
\end{align}
\end{subequations}
The new coefficients can be expressed in terms
of the gluing maps $\{h_I(z)\}$ as follows \cite{peskin2}
\begin{subequations}
\begin{align}
\Nc_{00}^{IJ}&=\bigl(1-\delta^{IJ}\bigr)\,\log
\Biggl[\frac{|h_I(0)-h_J(0)|}{
|h'_I(0)h'_J(0)|^{\frac12}}\Biggr],
\label{Nc00}
\\
\Nc^{IJ}_0(z)&\equiv\sum_{n=1}^{\infty}
\Nc_{0n}^{IJ}\,\sqrt{n}\,z^{n-1}
\notag
\\ =-&\frac{h'_J(z)}{h_I(0)-h_J(z)}-\frac{\delta^{IJ}}{z}
-\frac{h_J''(z)}{2h_J'(z)},
\label{Ncm0}
\\
\Nc^{IJ}(z,z')&\equiv
\sum_{m,n=1}^{\infty}\Nc_{mn}^{IJ}\sqrt{mn}\,z^{m-1}\,z^{\prime\,n-1}
\notag\\
=&\frac{h'_I(z)h'_J(z')}{\bigl[h_I(z)-h_J(z')\bigr]^2}
-\frac{\delta^{IJ}}{(z-z')^2}.
\label{Ncmn}
\end{align}
\label{Nc's}
\end{subequations}
Notice that all functions here are manifestly
$PSL(2)$ invariant \cite{peskin2} [pp. 487--488], i.e. they do not change under
$\{h_I\}\to\{T\circ h_I\}$ with $T(z)\in PSL(2)$.
Therefore all Neumann coefficients $\Nc_{mn}$ are $SL(2,\Rh)$ invariant.
Let us remember that, for example, the generating function
for the coefficients $N_{0n}^{IJ}$ \textit{does} depend on a choice
of the $SL(2,\Rh)$ frame. However this dependence is cancelled by
the nonanomalous momentum conservation, and therefore the $X$
vertex is $SL(2)$ invariant.

Writing the vertex in the form \eqref{expNc} eliminates the explicit
dependence on the background charge $Q$. In this notation
the $X^{\mu}$ vertex can be obtained simply by changing $\varepsilon$ to
$-\eta^{\mu\nu}$, $j_0=-i\sqrt{2\ap}p^{\mu}\,$
and $a^{\pm}_n\to\mp i\, \frac{\alpha_{\mp_n}^{\mu}}{\sqrt{n}}$.
Of course some of the terms in Eq. \eqref{Ncm0}
will drop out due to the normal (nonanomalous) momentum conservation \cite{peskin2}.

For Witten's $N$-string vertex the maps $h_I(z)$ are
\begin{subequations}
\begin{align}
h_I(z)&=\left(\frac{1-iz}{1+iz}\right)^{2/N}=e^{i\varphi_I} e^{4w/N},
\\
h_I'(z)&=-\frac{4i}{N}\,\bigl(\cosh w\bigr)^2 h_I(z),
\end{align}
\label{h_I}
\end{subequations}
where $z=i\tanh w$ and $\varphi_I=\frac{2\pi}{N}(\alpha_N-I)$.
Here $\alpha_N$ is a real number which is chosen in such a way
that all angles $\varphi_I$ lie in the range $(-\pi,\pi]$.

\subsection{Diagonalization of the Neumann Coefficients}
\subsubsection{Operator $\Nc^{IJ}$}
The operator $\Nc^{IJ}(z,z')$ was diagonalized
in Section~III of \cite{Belov:2003df}:
\begin{multline}
\Nc^{IJ}(z,-\oz')=\int_{-\infty}^{\infty}d\kappa\,
\mu_{1,N}^{IJ}(\kappa)
\\
\times|\kappa,1\rangle(z)\otimes\langle\kappa,1|(\oz'),
\label{Nco}
\end{multline}
where $|\kappa,1\rangle(z)$ is defined in \eqref{kappas},
$\langle\kappa,1|(\oz')\equiv\overline{|\kappa,1\rangle(z')}\;$ and
\begin{subequations}
\begin{align}
\mu_{1,N}^{II}(\kappa)&=-\frac{\sinh(N-2)\rx}{\sinh N\rx},
\\
\mu_{1,N}^{IJ}(\kappa)&=e^{+\rx (N+2I-2J)}\,\frac{\sinh 2\rx}{\sinh N\rx}
\quad(I<J),
\\
\mu_{1,N}^{IJ}(\kappa)&=e^{-\rx (N-2I+2J)}\,\frac{\sinh 2\rx}{\sinh N\rx}
\quad (I>J).
\end{align}
\label{mu1N}
\end{subequations}
Here $\rx\equiv\frac{\pi\kappa}{4}$.

\subsubsection{Matrix $\Nc^{IJ}_{00}$}
Substitution of the maps \eqref{h_I} into \eqref{Nc00} yields
\begin{equation}
\Nc_{00}^{IJ}=(1-\delta^{IJ})\,\log\left[
\frac{N}{2}\,\sin\left(\frac{\pi}{N}\,|I-J|\right)
\right].
\label{n00}
\end{equation}
This expression coincides with the matrix $M^{\prime\,IJ}_{N,\,00}$
which is defined by equation (5.8) in \cite{Belov:2003df}.
The numbers $\Nc_{00}^{IJ}$ can also be represented by
the following integral
\begin{multline}
\Nc_{00}^{IJ}=-\int_{-\infty}^{\infty}d\kappa\,
\frac{A_1(\kappa)}{\kappa^2}\,
\biggl\{
\frac12\,\Bigl[\mu^{IJ}_{1,N}(\kappa)+\mu^{JI}_{1,N}(\kappa)\Bigr]
\\
-\mu^{II}_{1,N}(\kappa)+\delta^{IJ}-1
\biggr\}.
\label{Nc00kappa}
\end{multline}
The simplest way to obtain this expression is to notice that
it can also be written as the following limit
\begin{multline*}
-\lim_{s\to +0}\int_{-\infty}^{\infty}d\kappa\,
A_s(\kappa)\,
\biggl\{
\frac12\,\Bigl[\mu^{IJ}_{s,N}(\kappa)+\mu^{JI}_{s,N}(\kappa)\Bigr]
\\
-\mu^{II}_{s,N}(\kappa)+\delta^{IJ}-1
\biggr\},
\end{multline*}
which was calculated in Eqs. (5.7) -- (5.9)
in \cite{Belov:2003df} with the result $M^{\prime\,IJ}_{N,\,00}$.

\subsubsection{Vector $\Nc_{0}^{IJ}$}
\begin{figure}[t]
\centering
\includegraphics[width=220pt]{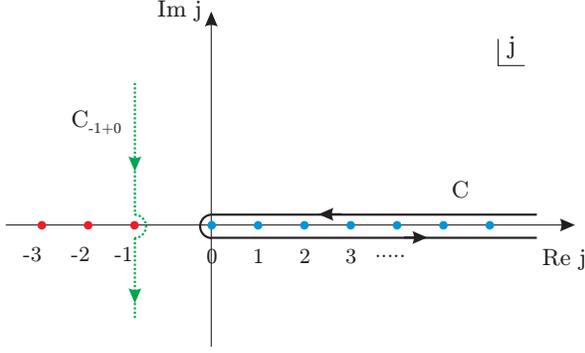}
\caption{The dots represent the poles
of the integrand in \eqref{gggg}. Contour
$C$ encircles the positive real
axis counterclockwise. Then we deform it to
the contour $C_{-1+0}$, which lies parallel to the
imaginary axis at $\Re j =-1$ and passes the pole at $j=-1$
on the right.}
\label{fig:2}
\end{figure}

Substitution of the maps \eqref{h_I} into \eqref{Ncm0}
yields
\begin{multline}
\Nc_0^{IJ}(-\oz')=\frac{4i}{N}\,
\bigl(\cosh \ow'\bigr)^2\,
\frac{e^{i(\varphi_J-\varphi_I)} e^{4\ow'\!/N}}{1-e^{i(\varphi_J-\varphi_I)} e^{4\ow'\!/N}}
\\
+\frac{\delta^{IJ}}{\oz'}-\frac{\oz'}{1+\oz^{\prime\,2}}+\frac{2i}{N}\,\bigl(\cosh \ow'\bigr)^2.
\label{Nc0-I}
\end{multline}
Assuming $\Re \ow'<0$ we can expand the first term in a binomial series,
and then rewrite it as a contour integral. All other terms
we leave as they are for a while. So the first term becomes
\begin{multline}
\text{first term}=-\frac{4i}{N}\,
\bigl(\cosh \ow'\bigr)^2\,
\oint_{C}\frac{dj}{2i\sin(\pi j)}\,
\\
\times
e^{i(\varphi_J-\varphi_I\pm\pi)(j+1)}
e^{\frac{4\ow'}{N}(j+1)},
\label{gggg}
\end{multline}
where the contour $C$ encircles the positive real axis in
the counterclockwise direction (see Figure~\ref{fig:2}).
Notice the sign ambiguity in the exponential, which comes
from the analytic continuation of $(-1)^j$.

Now we are going to deform the contour $C$ to lie parallel to
the imaginary axis as in Figure~\ref{fig:2}. To this end we need to worry about
the convergence as $|\Im j|\to\infty$. Let us assume $I\leqslant J$.
In this case the integral will have the exponential falloff if we
choose the upper ``$+$'' sign and temporary assume $0<-\Im \ow'<\frac{\pi}{4}$.
With these assumptions we can safely deform the contour $C$
as in Figure~\ref{fig:2} by
writing
\begin{equation*}
j=-1+0-\frac{iN\kappa}{4}
\end{equation*}
to get
\begin{multline}
\text{first term}=\bigl(\cosh \ow'\bigr)^2
\int_{-\infty}^{\infty}\frac{d\kappa}{2\sinh\bigl(\frac{\pi N\kappa}{4}+i0\bigr)}\,
\\
\times
e^{\frac{\pi\kappa}{4}(N+2I-2J)}\,e^{-i\kappa \ow'}.
\label{Nc0-II}
\end{multline}
Using the fact that
\begin{equation*}
\frac{1}{\sinh\bigl(\frac{\pi N\kappa}{4}+i0\bigr)}=
\mathscr{P}\frac{1}{\sinh\frac{\pi N\kappa}{4}}-\frac{4i}{N}\,\delta(\kappa)
\end{equation*}
one finds that the term in Eq. \eqref{Nc0-II} coming from the $\delta$-function
cancels the last term in Eq. \eqref{Nc0-I}.
Now we represent $1/\oz'$ in Eq. \eqref{Nc0-I} as in Eq.
\eqref{II-2} with the ``$+$'' sign, and write $\oz'/(1+\oz^{\prime\,2})$
as in Eq. \eqref{mid2} to get
\begin{multline}
\Nc_0^{IJ}(-\oz')=\int_{-\infty}^{\infty}d\kappa\,
\mathscr{P}\frac{\sqrt{A_1(\kappa)}}{\kappa}\,
\\
\times
\Bigl(\mu_{1,N}^{IJ}(\kappa)-1+\delta^{IJ}\Bigr)
\,\langle\kappa,1|(\oz'),
\label{Nc0-final}
\end{multline}
where $\mu_{1,N}^{IJ}(\kappa)$ is defined in \eqref{mu1N}.
Notice now that the integral in the rhs converges for all $\ow'$ in the
strip $|\Im \ow'|<\frac{\pi}{4}$. Therefore (by the standard
analytic continuation arguments) this expression
represents the vector $\Nc_0^{IJ}(z)$ for all $z$ in the unit disk.
One can easily check that equation \eqref{Nc0-final}
remains the same if we assume $I\geqslant J$ and choose
the ``$-$'' sign in Eqs. \eqref{Nc0-I} and \eqref{II-2}.

\subsection{The $N$-string Vertex in the Continuous
Basis}

\subsubsection{Vertex in the continuous basis}
We introduce the $s=1$ continuum oscillators
\begin{multline}
a^{\pm}(\kappa)=\sum_{n=0}^{\infty}a^{\pm}_{n+1}\,\langle\kappa,1|n,1\rangle
\\
\text{with}\quad
[a^-(\kappa),\,a^+(\kappa')]=-\varepsilon \delta(\kappa-\kappa').
\label{a(k)}
\end{multline}
In order to write the vertex we also need a twist operator $C$,
which roughly speaking corresponds to the substitution $\oz'\to -\oz'$.
It is defined \cite{GJ} in the discrete basis by $(C a^{\pm})_n=(-1)^n a^{\pm}_n$.
Hence in the continuous basis it becomes
\begin{equation}
\bigl(C a^{\pm}\bigr)(\kappa)=-a^{\pm}(-\kappa).
\end{equation}
Finally, substituting Eqs. \eqref{Nco}, \eqref{Nc0-final}
into \eqref{expNc} and using \eqref{a(k)} one obtains
\begin{widetext}
\begin{multline}
\langle V_{N,\,Q}^{(0)}|=
\sum_{\{q^I\}}\bigotimes_{I=1}^{N}\,{}_I\langle-q^I-Q|\,
\delta_{j^1_0+\dots+j^N_0+Q,0}\,
\exp\Biggl[
-\frac{\varepsilon}{2}\sum_{I,J=1}^{N}\,\int_{-\infty}^{\infty}
d\kappa\,a^{-(I)}(\kappa)\,\mu^{IJ}_{1,N}(\kappa)\,
\bigl(C a^{-(J)}\bigr)(\kappa)
\\
+\varepsilon \sum_{I,J=1}^{N}j_0^I
\int_{-\infty}^{\infty}d\kappa\,
\mathscr{P}\frac{\sqrt{A_1(\kappa)}}{\kappa}\,
\Bigl\{\mu^{IJ}_{1,N}(\kappa)-1+\delta^{IJ}\Bigr\}\bigl(C a^{-(J)}\bigr)(\kappa)
+\frac{\varepsilon}{2}\sum_{I,J=1}^{N}\,j_0^I\Nc^{IJ}_{00} j_0^J\Biggr],
\label{VNc}
\end{multline}
where $j_0$ is the ghost number operator,
$\langle -q-Q|j_0=q\langle -q-Q|$,
the continuous oscillators $a^-(\kappa)$ are
defined  in \eqref{a(k)},
the $N$-string Neumann matrix eigenvalues $\mu_{1,N}^{IJ}$
are defined in \eqref{mu1N}, and the coefficients $\Nc^{IJ}_{00}$
are listed in \eqref{n00}.
Notice that the gluing vertex for the matter field $X^{\mu}$ can be
obtained from this by putting $Q=0$ and replacing $\varepsilon$
with $-\eta^{\mu\nu}$, $j_0=-i\sqrt{2\ap} p_{\mu}$
and $a^{-(I)}\to a^{-(I)}_{\mu}$ (compare \eqref{VNc}
with Eq. (5.12) from \cite{Belov:2003df}).
\end{widetext}

\subsubsection{The unitary transformation}
In \cite{Belov:2003df} it was shown that the $s=0$ $N$-string matter
vertex can be obtained from the zero momentum (i.e. $s=1$) vertex
by the unitary transformation
\begin{multline}
U_p=\exp\biggl\{i\sqrt{2\ap}\,\hat{p}\int_{-\infty}^{\infty}d\kappa\,
\mathscr{P}\frac{\sqrt{A_1(\kappa)}}{\kappa}
\\
\times
\Bigl[a^+(\kappa)+a^-(\kappa)\Bigr]
\biggr\}.
\label{U_p}
\end{multline}
In the case of nonzero background charge $Q$ the
appropriate unitary operator is
\begin{multline}
U_{j_0}=\exp\biggl\{\varepsilon\Bigl(j_0+\frac{Q}{2}\Bigr)\int_{-\infty}^{\infty}d\kappa\,
\mathscr{P}\frac{\sqrt{A_1(\kappa)}}{\kappa}
\\
\times
\Bigl[a^+(\kappa)+a^-(\kappa)\Bigr]
\biggr\}.
\label{U_j0}
\end{multline}
Here the linear combination of $j_0$ and $Q$ was fixed
from the conjugation property \cite{fms} of $j_0$: $j_0^{\dag}=-j_0-Q$.
Notice that in contrast to \eqref{U_p}
the unitary transformation \eqref{U_j0} remains non-trivial
even for $j_0=0$.
Under this unitary transformation the $s=1$
continuum oscillator $a^{\pm}(\kappa)$
transforms to the $s=0$ oscillator
\begin{multline}
a^{\pm}(\kappa,j_0)\equiv U^{-1}_{j_0}a^{\pm}(\kappa)U_{j_0}
\\
=\pm\Bigl(j_0+\frac{Q}{2}\Bigr)\,\mathscr{P}\frac{\sqrt{A_1(\kappa)}}{\kappa}
+a^{\pm}(\kappa).
\end{multline}

Now let us try to proceed as in \cite{Belov:2003df} and
add the ghost numbers by applying
$N$ copies of the unitary transformation \eqref{U_j0}.
To this end we have to regularize the principal value in \eqref{U_j0}.
It does not matter what regularization we choose,
the final result should be regularization independent.
For definite we will assume the following
regularization \cite{Belov:2003df}
\begin{multline}
U_{j_0,s}=\exp\biggl\{\varepsilon\Bigl(j_0+\frac{Q}{2}\Bigr)
\,\int_{-\infty}^{\infty}d\kappa\,
\xi_s(\kappa)
\\
\times
\Bigl[a^+(\kappa)+a^-(\kappa)\Bigr]
\biggr\},
\label{U_j0r}
\end{multline}
where
\begin{equation}
\xi_s(\kappa)=\sqrt{A_s(\kappa)}=\frac{\sgn(\kappa)}{\sqrt{\kappa^2+4s^2}}\,
\sqrt{A_{1+s}(\kappa)}.
\end{equation}
Now $U_{j_0}$ can be normal ordered
\begin{equation}
U_{j_0,s}=\exp\left\{-\frac{\varepsilon}{2}\Bigl(j_0+\frac{Q}{2}\Bigr)^2\,\langle\xi_s,\xi_s\rangle
\right\}\,
:U_{j_0,s}\!:,
\label{Ureg}
\end{equation}
where
\begin{equation*}
\langle\xi_s,\xi_s\rangle=\int_{-\infty}^{\infty}d\kappa\,
A_s(\kappa)=\Gamma(2s).
\end{equation*}
Whenever possible we will write $\langle\xi_s,\xi_s\rangle$
instead of its value $\Gamma(2s)$. We will do this in order to be able to
choose another regularization without extra problems.

So we want to calculate
\begin{equation}
\lim_{s\to+0}\langle V^{(1)}_{N}|\bigotimes_{I=1}^N U_{j_0^{I},s}\,\delta_{j_0^1+\dots+j_0^N+Q,0},
\end{equation}
where the vertex $\langle V^{(1)}_{N}|$ is defined by the
first line in \eqref{VNc}. Substitution of Eq. \eqref{Ureg} yields
\begin{widetext}
\begin{multline*}
\lim_{s\to +0}
\sum_{q^1,\dots,q^N}\bigotimes_{I=1}^N{}_I\langle -q^I-Q|\,
\delta_{q^1+\dots+q^N+Q,0}
\exp\biggl\{
-\frac{\varepsilon}{2}\,\sum_{I,J=1}^{N}\int_{-\infty}^{\infty}
d\kappa\,a^{-(I)}(\kappa)\,\mu^{IJ}_{1,N}(\kappa)
\bigl(C a^{-(J)}\bigr)(\kappa)
\\
+\varepsilon
\sum_{I,J=1}^{N} \Bigl(q^I+\frac{Q}{2}\Bigr)\int_{-\infty}^{\infty}
d\kappa\,\xi_s(\kappa)\,\Bigl[\mu^{IJ}_{1,N}(\kappa)
+\delta^{IJ}
\Bigr]
\bigl(C a^{-(J)}\bigr)(\kappa)
\\
-\frac{\varepsilon}{2}
\sum_{I,J=1}^{N} \Bigl(q^I+\frac{Q}{2}\Bigr)\Bigl(q^J+\frac{Q}{2}\Bigr)
\int_{-\infty}^{\infty}
d\kappa\,\xi_s(\kappa)
\Bigl[\mu^{IJ}_{1,N}(\kappa)
+\delta^{IJ}
\Bigr]
\xi_s(\kappa)
\biggr\}
\end{multline*}
One sees that the integrals in the first and second term in the exponential
are well defined as $s\to 0$, but there is a problem with the $s\to 0$
limit in the third term.
\end{widetext}
In the second term we can substitute $Q=-\sum_I q^I$ and
use that $\sum_I [\mu^{IJ}(\kappa)+\delta^{IJ}]=2$ to obtain
\begin{multline*}
\text{second term}=
\varepsilon
\sum_{I,J=1}^{N} q^I\int_{-\infty}^{\infty}
d\kappa\,\mathscr{P}\frac{\sqrt{A_1(\kappa)}}{\kappa}
\\
\times\Bigl[\mu^{IJ}_{1,N}(\kappa)
+\delta^{IJ}-1
\Bigr]
\bigl(C a^{-(J)}\bigr)(\kappa).
\end{multline*}
In the third term, we can use equation \eqref{Nc00kappa} and the anomalous conservation
law to get
\begin{multline*}
\text{third term}=\frac{\varepsilon}{2}\sum_{I,J=1}^N q^I\Nc_{00}^{IJ}q^J
\\
-\frac{\varepsilon\,Q^2}{2}\,\int_{-\infty}^{\infty}d\kappa\,
\frac{A_1(\kappa)}{\kappa^2}\,\Bigl[\mu^{II}_{1,N}(\kappa)-\mu_{1,N}^{II}(0)\Bigr]
\\
-\frac{\varepsilon\,Q^2}{2}\,\frac{(N-2)^2}{2N}\,\langle\xi_s,\xi_s\rangle.
\end{multline*}
The integral here is easy to calculate, and we finally get the relation
\begin{multline}
\langle V_{N,Q}^{(0)}|\{q^I\}\rangle
=\lim_{s\to+ 0}
e^{F_{N,s}}\,
\langle V_{N,Q}^{(1)}|
\\
\times U_{q^1}\otimes
\dots
\otimes U_{q^N}\,\delta_{q^1+\dots+q^N+Q,0}\, ,
\label{VNQU}
\end{multline}
where
\begin{multline}
F_{N,s}\equiv
\frac{\varepsilon\,Q^2}{2}\,\frac{(N-2)^2}{2N}\,\langle\xi_s,\xi_s\rangle
\\
+\frac{\varepsilon\,Q^2}{2}\,\Bigl[\log\frac{N}{2}-\frac{N-2}{N}\,\log 2\Bigr].
\label{beta_Ns}
\end{multline}
Notice that for general $Q$ the function $F_{N,s}$ is zero only for $N=2$.
This means that
the string inner product $\langle V_2^{(0)}|$ is not affected by this singularity.
For $Q=0$ $F_{N,s}$ is identically zero, and hence
after replacing $q^I\to -i\sqrt{2\ap} p^I_{\mu}$
equation \eqref{VNQU} reproduces the result of \cite{Belov:2003df}
for the matter sector.

\section{Associativity of Witten's Star Product}
\label{sec:assoc}
\setcounter{equation}{0}
\subsection{Descent Relation between the Gluing Vertices}
The aim of this section is to verify the following descent relation:
\begin{equation}
{}_{1\dots N}\langle V_{N}|={}_{1\dots N,N+1}\langle V_{N+1}|
V_1\rangle_{N+1},
\label{descent}
\end{equation}
where $\langle V_N|$ is the combined matter+ghost gluing
$N$-string vertex. The combined vertex $\langle V_N|$
has the form \eqref{VNc} with the replacements:
$a^{-}\to a^-_{\mu}$ ($\mu=-1,\dots,D-1$) where
$\mu=-1$ corresponds to the ghost oscillator and
$\mu=0,\dots,D-1$ to the matter ones; $\varepsilon \to-\eta^{\mu\nu}$
where $\eta^{\mu\nu}=\mathrm{diag}(-\varepsilon,-1,1,\dots,1)$;
and $j_0\to j_0^{\mu}=(j_0,-i\sqrt{2\ap}\vec{p}\,)$.

We know that the vertex depending on the momenta and ghost numbers
can be obtained from the $s=1$ vertex by the unitary transformation
as in \eqref{VNQU}. Since adding the momentum
does not produce any divergencies \cite{Belov:2003df} we put it
equal to zero.
So we need to calculate the following product
\begin{multline}
e^{F_{N+1,s}+F_{1,s}}\,
{}_{1\dots N,N+1}\langle V_{N+1}^{(1)}
|V_{1}^{(1)}\rangle_{N+1}
\\
\times U_{q^1}\otimes
\dots
\otimes U_{q^N}\,\delta_{q^1+\dots+q^N+Q,0}.
\label{D1}
\end{multline}
The inner product in the $(N+1)^{\text{th}}$ tensor component is easy to calculate:
\begin{widetext}
\begin{multline*}
{}_{1\dots N,N+1}\langle V_{N+1}^{(1)}
|V_{1}^{(1)}\rangle_{N+1}
=\det\bigl(1-\mu_{N+1}^{11}\bigr)^{-\frac{D+1}{2}}\,
\\
\times{}_{1\dots N}\langle 0|\exp\left[
\frac{1}{2}\,\sum_{\mu,\nu=-1}^{D-1}\sum_{I,J=1}^{N}\,
\int_{-\infty}^{\infty}d\kappa\,
a^{-(I)}_{\mu}(\kappa)\,\eta^{\mu\nu}
\left\{\mu_{N+1}^{IJ}+
\frac{\mu_{N+1}^{I,N+1}\mu_{N+1}^{N+1,J}}{1-
\mu_{N+1}^{11}}\right\}(\kappa)\,
\bigl(C a^{-(J)}_{\nu}\bigr)(\kappa)\right],
\end{multline*}
\end{widetext}
where $\mu_{N}^{IJ}\equiv \mu_{1,N}^{IJ}$.
It is a matter of simple algebra to show that
the term in $\{\}$ equals $\mu^{IJ}_{N}(\kappa)$.
Hence we see that the descent relation is actually satisfied
up to a numeric coefficient
\begin{equation}
{}_{1\dots N,N+1}\langle V_{N+1}|V_1\rangle_{N+1}=
(\Zc_{N+1,1;N})\,{}_{1\dots N}\,\langle V_{N}|
,
\label{true_descent}
\end{equation}
where
\begin{multline}
\log\Zc_{N+1,1;N}\equiv -\frac{D+1}{2}\,\log \det\bigl(1-\mu_{N+1}^{11}\bigr)
\\
+F_{N+1,s}+F_{1,s}-F_{N,s}
\end{multline}
and $F_{N,s}$ is defined in \eqref{beta_Ns}.
In other words $\log\Zc_{N+1,1;N}$ is
\begin{multline}
\log\Zc_{N+1,1;N}\equiv
+\frac{\varepsilon Q^2}{2}\,\log\frac{N+1}{2N}
\\
-\frac{D+1}{2}
\int_{-\infty}^{\infty}
d\kappa\,\log\bigl(1-\mu_{N+1}^{11}(\kappa)\bigr)\,
\rho_{1}(\kappa)
\\
+\frac{\varepsilon Q^2}{2}\,
\frac{(N-1)(N+2)}{N(N+1)}\,\Bigl\{\langle \xi_s,\xi_s\rangle
+\log 2
\Bigr\},
\label{Z1N}
\end{multline}
where $\rho_1(\kappa)$ is the trace density which
was calculated in \cite{density}
\begin{multline*}
\rho_1(\kappa)=\frac{1}{\pi}\,\Bigl[\langle\xi_s,\xi_s\rangle+\log 2\Bigr]
\\
-\frac{1}{4\pi}
\biggl[\psi\Bigl(1+\frac{i\kappa}{2}\Bigr)+\psi\Bigl(1-\frac{i\kappa}{2}\Bigr)
+2\gamma_E\biggr].
\end{multline*}
From this it follows that $\Zc_{2,1;1}=1$.
If the descent relation \eqref{descent} were true,
all $\Zc_{N+1,1;N}$ with $N>1$ would be $1$. But as we will see in a moment
 $\Zc_{N+1,1;N}$ is a nontrivial function
of $N$, and therefore the vertices must contain an additional normalization
factor.

For the critical bosonic string ($D=26$, $\varepsilon=1$ and $Q=-3$)
$\Zc_{N+1,1;N}$ is a finite function of $N$
\begin{multline}
\log \Zc_{N+1,1;N}=
-\frac{9}{2}\,\sum_{\alpha\in A}s_{\alpha}\log\alpha
\\
+\frac{27}{2}\,
\sum_{\alpha\in A}s_{\alpha} D_0(\alpha),
\label{logZ1N}
\end{multline}
where $A=\{2,N,N+1,1\}$, $\{s_{\alpha}|\alpha\in A\}=\{1,1,-1,-1\}$ and
\begin{equation}
D_0(\alpha)=\int_0^{\infty}\frac{dt}{t}\,
\left\{
\frac{\coth \frac{t}{\alpha}-\frac{\alpha}{t}}{e^{t}-1}
-\frac{1}{3\alpha}\,\frac{1}{1+t}
\right\}.
\label{D(a)}
\end{equation}
The details of this calculation will be presented in
\cite{BG}.
From the representation \eqref{logZ1N} it follows that
$\Zc_{N+1,1;N}$ can be written as
\begin{equation}
\Zc_{N+1,1;N}=\frac{\Zc_{N}}{\Zc_{1}\Zc_{N+1}}.
\label{Z1NZN}
\end{equation}
Here the logarithm of $\Zc_{N}$ is given by
\begin{equation}
\log \Zc_N=
-\frac{9}{2}\,\log\frac{N}{2}
+\frac{27}{2}\,\Bigl[D_0(N)-D_0(2)\Bigr].
\label{ZcN}
\end{equation}
Notice that the function $\Zc_N$
can \textit{not} be uniquely determined
from the relation \eqref{Z1NZN}.
It is defined up to a rescaling
\begin{equation}
\Zc_N\mapsto (\text{const})^{N-2}\Zc_N.
\label{Zrescaled}
\end{equation}
The function $\Zc_{N}$ defined in \eqref{ZcN}
monotonically goes to zero on the interval $[1,\infty)$,
and its asymptotic at infinity is
\begin{equation*}
\Zc_{N}\propto \left[\frac{2}{N}\right]^{9/2}\,
\exp\biggl\{-\frac{27}{2} D_0(2)+O\Bigl(\frac{1}{N}\Bigr)\biggr\}.
\end{equation*}

Now it is obvious that in order to have the descent
relation \eqref{descent} one
has to introduce the normalized gluing vertices $\la V_N|$,
which are defined by
\begin{equation}
\la V_{N}|=\Zc_{N}\,\langle V_{N}|
\quad\text{for}\quad N\geqslant 1.
\label{<<VN|}
\end{equation}
The normalization $\Zc_N$ is given by \eqref{ZcN}
or any of \eqref{Zrescaled}.
Notice that $\Zc_2=1$ independently on the choice of
the scaling factor in \eqref{Zrescaled}.
The ambiguity \eqref{Zrescaled} is closely
related to the string field redefinition.
Indeed the factor $(\text{const})^{N-2}$ in the vertex
can be cancelled by simultaneous rescaling of the string field
$\Ac\mapsto (\text{const})^{-1}\Ac$
and the coupling constant $g_{o}\mapsto (\text{const})^{-1}g_o$.
The natural choice of the factor in \eqref{Zrescaled}
is that where  $\Zc_N$
coincides with the partition function
of the bosonic matter+ghost CFT
on the gluing surface, which is the unit disk with Neumann boundary conditions and
the angle excess $\pi(N-2)$.
This choice is basically
follows from the relation
\begin{equation*}
\Zc_N=\la V_N|\bigl(|0\rangle_1\otimes\dots\otimes|0\rangle_N\bigr),
\end{equation*}
where $\la V_N|$ is supposed to be a surface (multi)state, and
therefore the rhs must be the partition function of this surface
[see Figure~5 in \cite{zwiebach}].

For $N=3$ the normalized as in Eq. \eqref{<<VN|} $3$-string
vertex coincides with Witten's original definition \cite{witten}.
In that paper he defined it as the Polyakov integral over
the gluing surface which, of course, includes the partition function
in its definition.

Notice that in Moyal formulation
of SFT (MSFT) \cite{Bars:2002nu} the
star product is associative by construction and
there is a way to obtain
the Neumann matrix elements from it \cite{Bars:2002nu}.
Therefore it should be also possible to extract the normalization
of the gluing vertices and compare them with $\Zc_N$.

\subsection{Associativity}
The associativity requires many relations between the gluing vertices.
For example,
\begin{multline}
\Bigl({}_{123}\langle V_3|\otimes {}_{456}\langle V_3|
\Bigr)|V_2\rangle_{34}= {}_{1256}\langle V_4|,
\\
\Bigl({}_{123}\langle V_3|\otimes {}_{456}\langle V_3|
\otimes {}_{78}\langle V_2|
\Bigr)|V_3\rangle_{368}= {}_{12457}\langle V_5|
\label{as_example}
\end{multline}
and many many more. The question is if these relations are satisfied
for the normalized vertices \eqref{<<VN|}.
Actually the question is only about the normalization factors,
since the exponentials were worked out in \cite{peskin2,
Furuuchi:2001df}.

We claim that the normalized vertices \eqref{<<VN|} indeed
satisfy the relations like \eqref{as_example}.
The proof is simple and does not require complicated calculations.
Let us prove, for example, the first relation in \eqref{as_example}.
Suppose that it is false, and there is a constant $A\ne 1$ in the rhs:
\begin{equation*}
\Bigl({}_{123}\la V_3|\otimes {}_{456}\la V_3|
\Bigr)|V_2\ra_{34}=A\; {}_{1256}\la V_4|.
\end{equation*}
Now we contract this equation with the identity state
$|V_1\ra{}_6$ in the $6$-th tensorial space. Using
the descent relations \eqref{descent} we obtain
\begin{equation*}
\Bigl({}_{123}\la V_3|\otimes {}_{45}\la V_2|
\Bigr)|V_2\ra_{34}=A\; {}_{125}\la V_3|.
\end{equation*}
Noticing that ${}_{45}\la V_2|V_2\ra_{34}=\delta_{53}$ one finds
that the constant $A$ equals $1$. So the first relation in \eqref{as_example}
is true.
Actually all relations like \eqref{as_example} can be proved in this
manner. Therefore the normalized vertices \eqref{<<VN|}
\textit{do} satisfy the associativity relations.

\section{Discussion}
\label{sec:disc}
Here I want to discuss some consequences of the normalization
\eqref{<<VN|} for the numeric calculations of the tachyon condensation
\cite{conj}.
First, the calculations in which one uses only vertices $\langle V_2|$
and $\langle V_3|$ (see for example \cite{tach}) are \textit{not}
affected by the normalization
\eqref{<<VN|}. One can simply cancel the factor
$\Zc_3$ in the cubic vertex by simultaneous rescaling of the string
field and the coupling constant as $\Ac\mapsto \Zc_3^{-1} \Ac$
and $g_o\mapsto \Zc_3^{-1} g_o$ correspondingly. However the calculations
in the bosonic string which involve the higher vertices
(if any were done)
have to be revised.

Second, the fact that $\Zc_N\ne 1$ ($N=1$ and $N\geqslant 3$)
for the bosonic string may potentially affect the numeric
calculations in the nonpolynomial fermionic string field theory
(see for example \cite{berkovits}). To check this
 one has to calculate the contribution
of the matter fermions and superghosts into the partition
function $\Zc_N$. For the same reason
as for the bosonic cubic SFT the calculations in the cubic fermionic string
field theory (see for example \cite{ABK}) are not affected.

\begin{acknowledgments}
I would like to thank B.~Doyon, P.~Fonseca and S.~Lukyanov for
helpful discussions.
I would like to thank A.~Giryavets, H.~Liu
and C.~Lovelace for many useful discussions and valuable comments on the draft
of this paper.
I would like to thank G.~Moore for support, discussion
and constant attention to my work.
I am very grateful to A.B.~Zamolodchikov for an illuminating discussion.
I would like to thank E.~Fuchs, M.~Kroyter
and A.~Konechny for interesting email correspondence.
The work was supported by DOE grant DE-FG02-96ER40959
and in part by RFBR grant 02-01-00695.
\end{acknowledgments}

\appendix
\renewcommand {\theequation}{\thesection.\arabic{equation}}

\section{Representation of $(z+\oz')^{-1}$ through $s=0\times s=1$
$K_1$-eigenfunctions}
\label{app:st}

In this Appendix we derive a representation of the
first term in Eq. \eqref{MII} through the tensor product of
$s=0$ and $s=1$ $K_1$-eigenfunctions \eqref{kappa0} and \eqref{kappas}.

We start from the following equation for
$0<\mp\Im(\ow'-w)\leqslant\frac{\pi}{2}$ and $s>0$
\begin{multline}
\frac{\Gamma(2s)}{(z+\oz')^{2s}}=
\bigl(\cosh w\cosh\ow'\bigr)^{2s}
\\
\times\int_{-\infty}^{\infty}d\kappa\,
A_s(\kappa)\,e^{\pm\frac{\pi\kappa}{2}}\,e^{i\kappa(w-\ow')}.
\label{zoz1}
\end{multline}
This expression follows from Eqs. (3.14) and (3.23) in \cite{Belov:2003df}.
Obviously differentiating the lhs with respect to $\oz'$ and
taking the limit $s\to 0$ one obtains $-(z+\oz')^{-1}$.
Hence the problem is to perform these operations on the rhs.

Differentiation by $\oz'=-i\tanh \ow'$ of \eqref{zoz1}'s rhs
yields
\begin{multline}
\bigl(\cosh w\bigr)^{2s} \bigl(\cosh \ow'\bigr)^{2s+2}\,\int_{-\infty}^{\infty}d\kappa\,
e^{\pm\frac{\pi\kappa}{2}}
\\
\times
A_s(\kappa)\,e^{i\kappa (w-\ow')}\bigl\{-2s\,\oz'+\kappa\bigr\}.
\label{zoz2}
\end{multline}
Using the following relations
\begin{multline*}
A_s(\kappa)=\frac{A_{1+s}(\kappa)}{\kappa^2+4s^2},
\quad
\lim_{s\to 0}\frac{\kappa}{\kappa^2+4s^2}=\mathscr{P}\frac{1}{\kappa},
\\
\text{and}\quad
\lim_{s\to 0}\frac{2s}{\kappa^2+4s^2}=\pi\delta(\kappa),
\end{multline*}
one can take the $s\to 0$ limit in Eq. \eqref{zoz2}:
\begin{equation*}
\bigl(\cosh \ow'\bigr)^{2}\,\int_{-\infty}^{\infty}d\kappa\,
e^{\pm\frac{\pi\kappa}{2}}\,
\mathscr{P}\frac{A_1(\kappa)}{\kappa}
\,e^{i\kappa (w-\ow')}
-\frac{\oz'}{1+\oz^{\prime\,2}}.
\end{equation*}
The last term in this expression comes from the mid-point
and therefore can be written as in Eq. \eqref{mid2}. Using Eqs. \eqref{kappa0}
and \eqref{kappas} we finally obtain
\begin{multline}
-\frac{1}{z+\oz'}=\int_{-\infty}^{\infty}d\kappa\,
e^{\pm\frac{\pi\kappa}{2}}\,|\kappa,0\rangle(z)\otimes\langle\kappa,1|(\oz')
\\
-\int_{-\infty}^{\infty}d\kappa\,
\mathscr{P}\frac{\sqrt{A_1(\kappa)}}{\kappa}\,
\langle\kappa,1|(\oz').
\end{multline}
Here the first integral converges for $0<\mp\Im(\ow'-w)\leqslant \frac{\pi}{2}$,
while the second integral converges for all $\oz'$ in the unit disk.

\newpage
{\small
\renewcommand{\baselinestretch}{1.1}

}

\end{document}